\documentclass[a4paper]{jpconf}

\usepackage{graphicx}
\usepackage{relsize}

\def\ra       {\!\rightarrow\!}
\def\cp       {$CP$}
\def\pt       {p^{}_T}

\def\dbar     {\overline{D}{}^0}
\def\bbar     {\overline{B}{}^0}

\def\babar{\mbox{\slshape B\kern-0.1em{\smaller A}\kern-0.1em
    B\kern-0.1em{\smaller A\kern-0.2em R}}}

\begin{document}

\title{\noindent
Experimental summary: step-by-step towards new physics \\
\vskip-2.2in
\hskip5.2in \normalsize{\rm UCHEP-16-08} 
\vskip2.0in
}

\author{A J Schwartz}

\address{
Physics Department, University of Cincinnati, P.O. Box 210011,
Cincinnati, Ohio 45221 USA }

\ead{alan.j.schwartz@uc.edu}

\begin{abstract}
We summarize some highlights from experimental results presented at 
the XIIth International Conference on Beauty, Charm, and Hyperons in 
Hadronic Interactions, held at George Mason University June 12-18, 2016.
\end{abstract}

\section{Introduction}

This year's workshop featured about fifty experimental
talks covering a wide variety of results in meson and baryon 
decays, heavy flavor and quarkonium production, heavy ion 
collisions, kaon physics, neutrino physics, and searches
for new physics (NP). The presentations were organized 
as follows:
heavy flavor production (Monday);
heavy flavor decays (Tuesday);
neutrino physics (Wednesday);
mixing and \cp\ violation (Thursday);
spectroscopy (Thursday); and
future experiments and facilities (Friday).
Interspersed throughout the week were talks
on searches for NP and theory talks. Experimental
results were presented from ATLAS, ALICE, Belle, 
BaBar, BESIII, CDF, CMS, LHCb, PHENIX, and STAR. 
In this summary I discuss only a few highlights from 
among all these talks. For further details the reader 
is referred to the original presentations and these Proceedings.

\section{Heavy flavor production}

Numerous results were presented on the production of 
$D$ and $B$ mesons, charmonium and bottomonium mesons,
$W^\pm$ and $Z^0$ vector bosons, and lighter
pions and electrons, from LHC and RHIC experiments. 

\begin{figure}
\begin{center}
\hbox{
\includegraphics[width=0.48\textwidth]{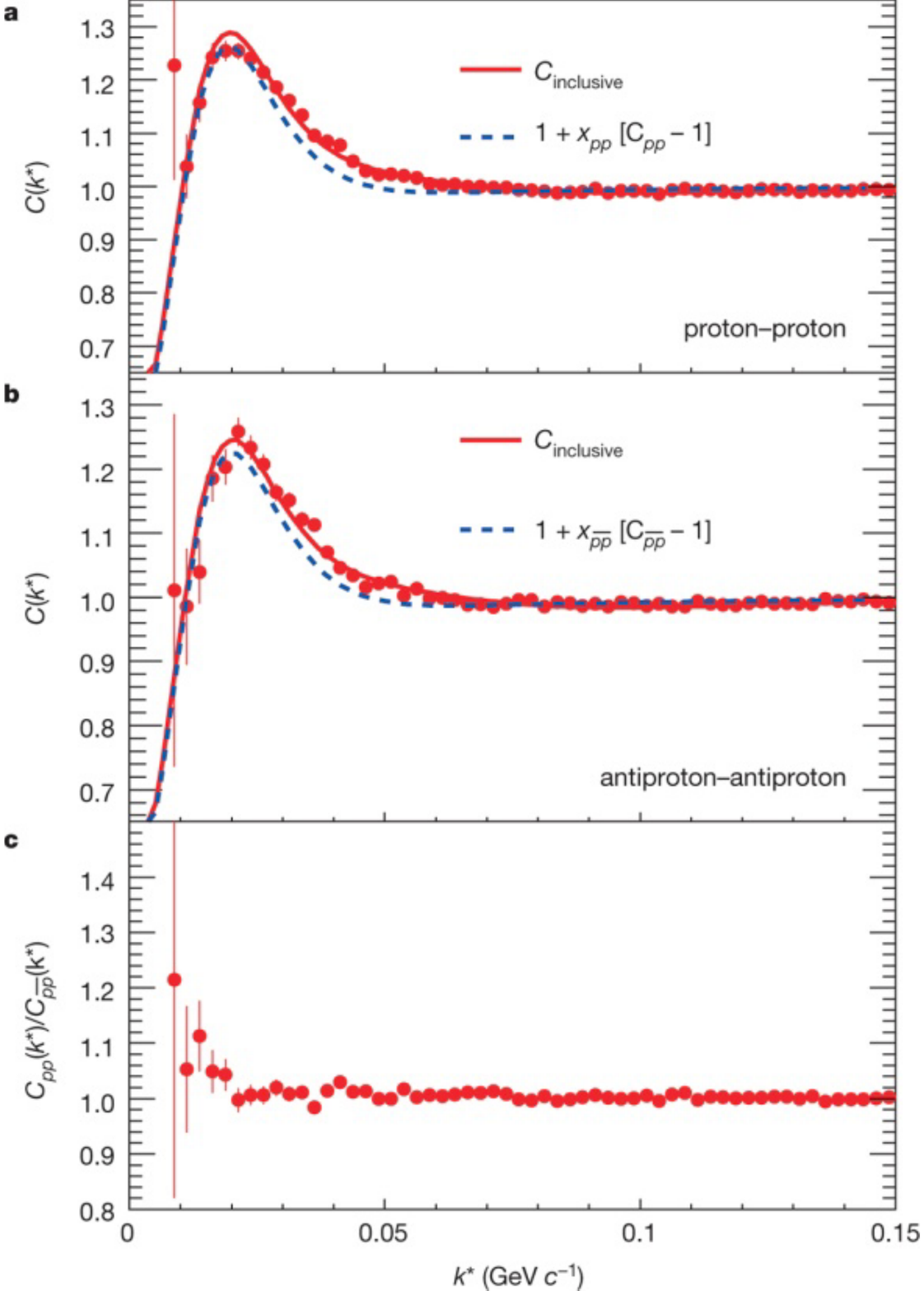}
\hskip0.15in
\includegraphics[width=0.45\textwidth]{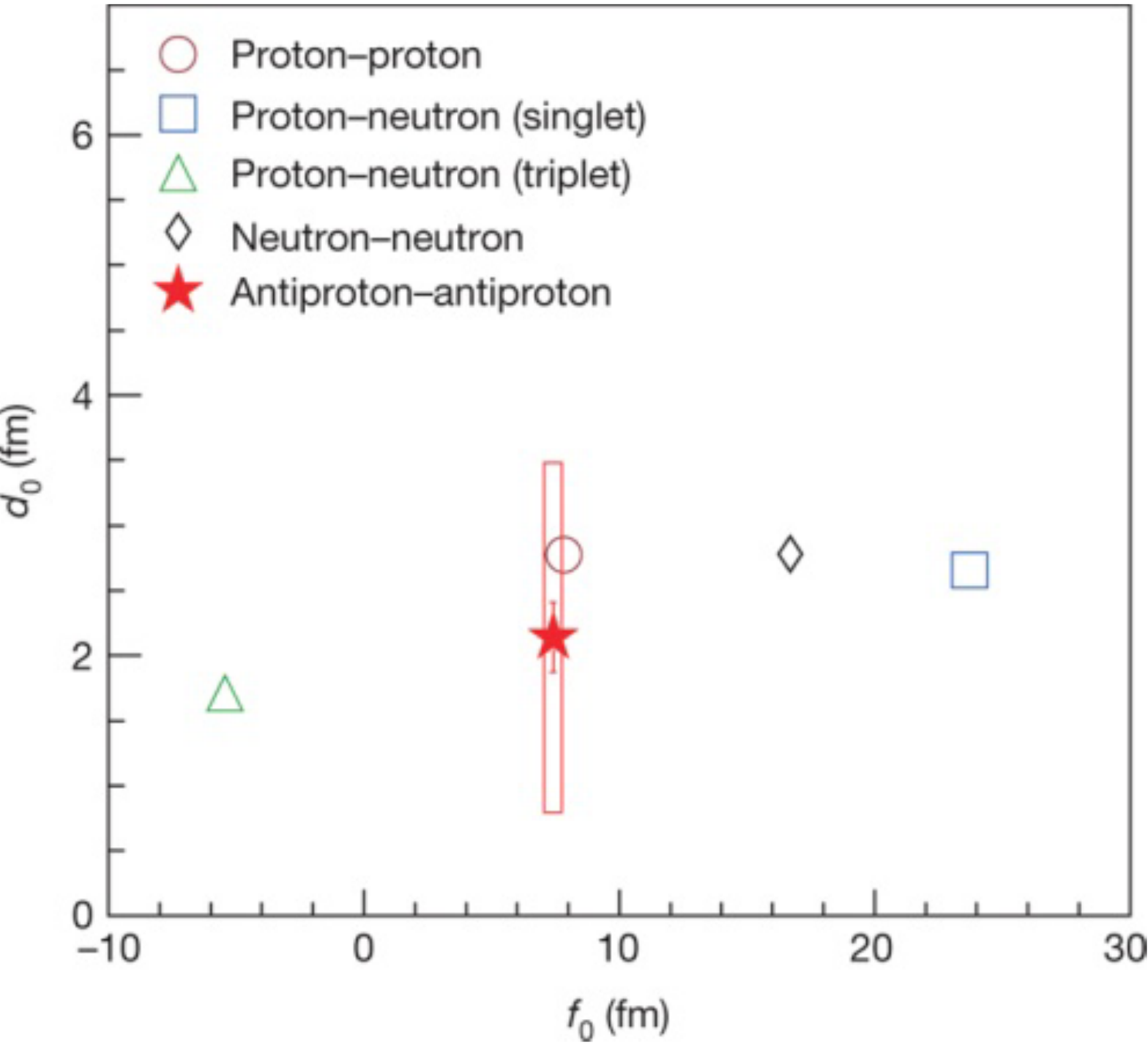}
}
\end{center}
\vskip-0.30in
\caption{\label{fig:star_corr}
Data from Au-Au collisions at STAR.
Left top: correlation function $C(k^*)^{}_{pp}$ for $pp$ tracks;
left middle: correlation function $C(k^*)^{}_{\bar{p}\bar{p}}$ 
for $\bar{p}\bar{p}$ tracks;
left bottom: ratio of $C(k^*)^{}_{pp}$ to $C(k^*)^{}_{\bar{p}\bar{p}}$.
The rise at low $k^*$ for the top and middle plots indicates an
attractive force.
Right: values of the range $d^{}_0$ and scattering length $f^{}_0$
resulting from fitting the correlation functions $C(k^*)^{}_{pp}$ and
$C(k^*)^{}_{\bar{p}\bar{p}}$. }
\end{figure}

The STAR experiment (Zhang) presented measurements~\cite{STAR_Nature} 
of the correlation function $C(k^*)\equiv A(k^*)/B(k^*)$, where
the variable $k^*$ is half the relative momentum between two particles
produced in a collision,
$A(k^*)$ is the distribution for a pair of particles in the same
event, and $B(k^*)$ is the distribution for two particles
produced in {\it different\/} events.
If there is a net attractive interaction between the particles, 
$C(k^*)$ increases as $k^*$ decreases; if there is a net repulsive 
interaction, $C(k^*)$ decreases as $k^*$ decreases; and if there is
negligible interaction, $C(k^*)$ is independent of $k^*$ and equals 
unity. The STAR data corresponds to Au-Au collisions at 
$\sqrt{s^{}_{NN}} = 200$~GeV and is shown in Fig.~\ref{fig:star_corr}(left).
A strong attractive correlation is observed for both $p$-$p$ and
$\bar{p}$-$\bar{p}$ distributions. Taking their ratio shows an 
equal attractive force over most of the $k^*$ range. However,
at the lowest $k^*$ values the $p$-$p$ attractive force appears
to be stronger. More data is needed to confirm this effect. 

To measure a quantitative difference between $p$-$p$ and $\bar{p}$-$\bar{p}$ 
interactions and in this manner test $CPT$, STAR fits the
$p$-$p$ and $\bar{p}$-$\bar{p}$ spectra for $C(k^*)$ using
the Lednicky and Lyuboshitz model~\cite{Lednicky}.
The fit yields two parameters: the effective range of the 
interaction $d^{}_0$, and the scattering length $f^{}_0$.
The latter indicates whether the interaction corresponds to
an attractive bound state ($f^{}_0<0$),
an attractive but unbound state ($f^{}_0>d^{}_0$),
or is repulsive ($0<f^{}_0<d^{}_0$). The fit result
is plotted in Fig.~\ref{fig:star_corr}(right), which shows
that $d^{}_0$ and also $f^{}_0$ are the same for $p$-$p$ 
and $\bar{p}$-$\bar{p}$ interactions, within errors.
In both cases the range $d^{}_0$ is approximately 2.5~fm, 
and $f^{}_0$ corresponds to an attractive unbound state.

\begin{figure}
\vskip-1.0in
\begin{center}
\includegraphics[width=0.6\textwidth]{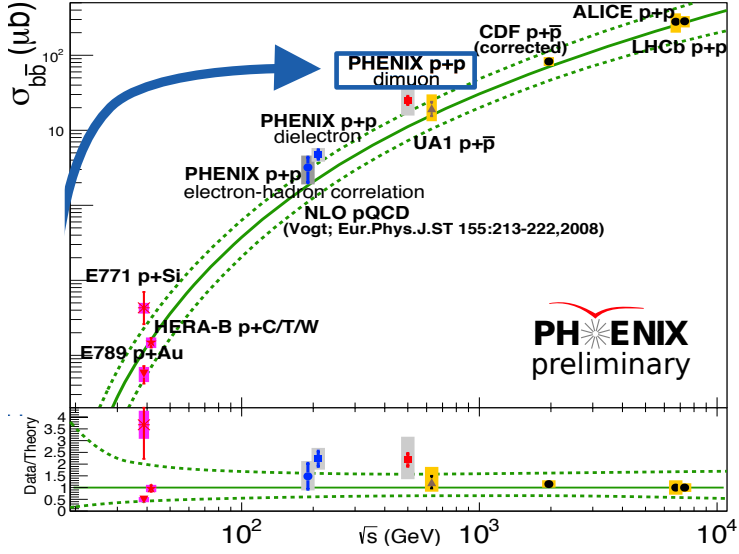}
\end{center}
\vskip-1.0in
\caption{
Measurements of the $b\bar{b}$ production cross section, and 
next-to-leading-order pQCD predictions~\cite{sigma_bb_theory}. }
\label{fig:sigma_bb}
\end{figure}

The LHCb experiment (Szumlak) presented measurements of
differential cross sections 
for prompt and secondary $J/\psi$ production, 
$\Upsilon(1S)$, $\Upsilon(2S)$, and $\Upsilon(3S)$ production,
and $D$ and $\Lambda^{}_b$ production. All cross sections
are measured as a function of $p^{}_T$ and rapidity $y$ 
and correspond to $pp$ collisions at $\sqrt{s}=7,8$~TeV.
From measurements of $D^0$, $D^{(*)+}$, and $D^+_s$ production,
LHCb obtains a $c\bar{c}$ total production cross section of
$(2940\pm 3\pm180\pm 160)$~$\mu$b, where the first error is
statistical, the second is systematic, and the last error is
due to the $c\ra D$ fragmentation model. From measurements
of non-prompt $J/\psi$ production, i.e., events in which the 
$J/\psi$ candidate forms a secondary (rather than primary) vertex, 
LHCb calculates a $b\bar{b}$ total production cross section of
$(515\pm 2.0\pm 53.0)$~$\mu$b. This value is plotted in 
Fig.~\ref{fig:sigma_bb} along with measurements of $\sigma_{b\bar{b}}$ 
presented by PHENIX (Haseler). The latter results
are from three independent analyses: 
one using same-sign $\mu^\pm\mu^\pm$ pairs;
one using opposite-sign $e^+e^-$ pairs; and 
one using electron-hadron correlations.
All $\sigma^{}_{b\bar{b}}$ measurements are in good
agreement with next-to-leading-order pQCD~\cite{sigma_bb_theory} 
over almost three orders of magnitude.

LHCb also measured the production of $B^+_c$ mesons by reconstructing
Cabibbo-favored $B^+_c\ra J/\psi\,\pi^+$ decays~\cite{charge-conjugates}. 
The signal yield is normalized to the number of
color-suppressed $B^+\ra J/\psi\,K^+$ decays reconstructed, and the
result is 
$R\equiv [\sigma^{}_{B^+_c}\cdot B(B^+_c\ra J/\psi\,\pi^+)]
/[\sigma^{}_{B^+}\cdot B(B^+\ra J/\psi\,K^+)] =
(0.683\pm 0.018\pm 0.009)\%$.
This corresponds to the kinematic range $2.5<y<4.5$ and $4<p^{}_T<20$~GeV/$c$.

The ALICE experiment (De) presented results for 
meson production in proton-nucleus
and nucleus-nucleus collisions. The energy density of such collisions
corresponds to the environment of a quark-gluon plasma (QGP), and partons
produced in collisions interact with the QGP when escaping and
subsequently lose energy. This parton energy loss is expected to
decrease as the parton mass increases. The parameter quantifying 
parton interactions with the QGP is the ``nuclear modification factor:''
\begin{eqnarray}
R_{AA}(\pt) & \equiv &  \frac{1}{N^{}_{\rm coll}} 
\frac{(dN^{}_{AA}/dp^{}_T)}{(dN^{}_{pp}/dp^{}_T)}\,,
\end{eqnarray}
where $N^{}_{\rm coll}$ is the number of nucleons participating
in the $A$-$A$ collision.
The greater the interaction with the QGP and subsequent energy loss,
the lower $R_{AA}(\pt)$; this is referred to as ``suppression.'' 
The amount of suppression is found to increase with $p^{}_T$.

\begin{figure}[htb]
\vskip-0.20in
\begin{center}
\hbox{
\includegraphics[width=0.32\textwidth]{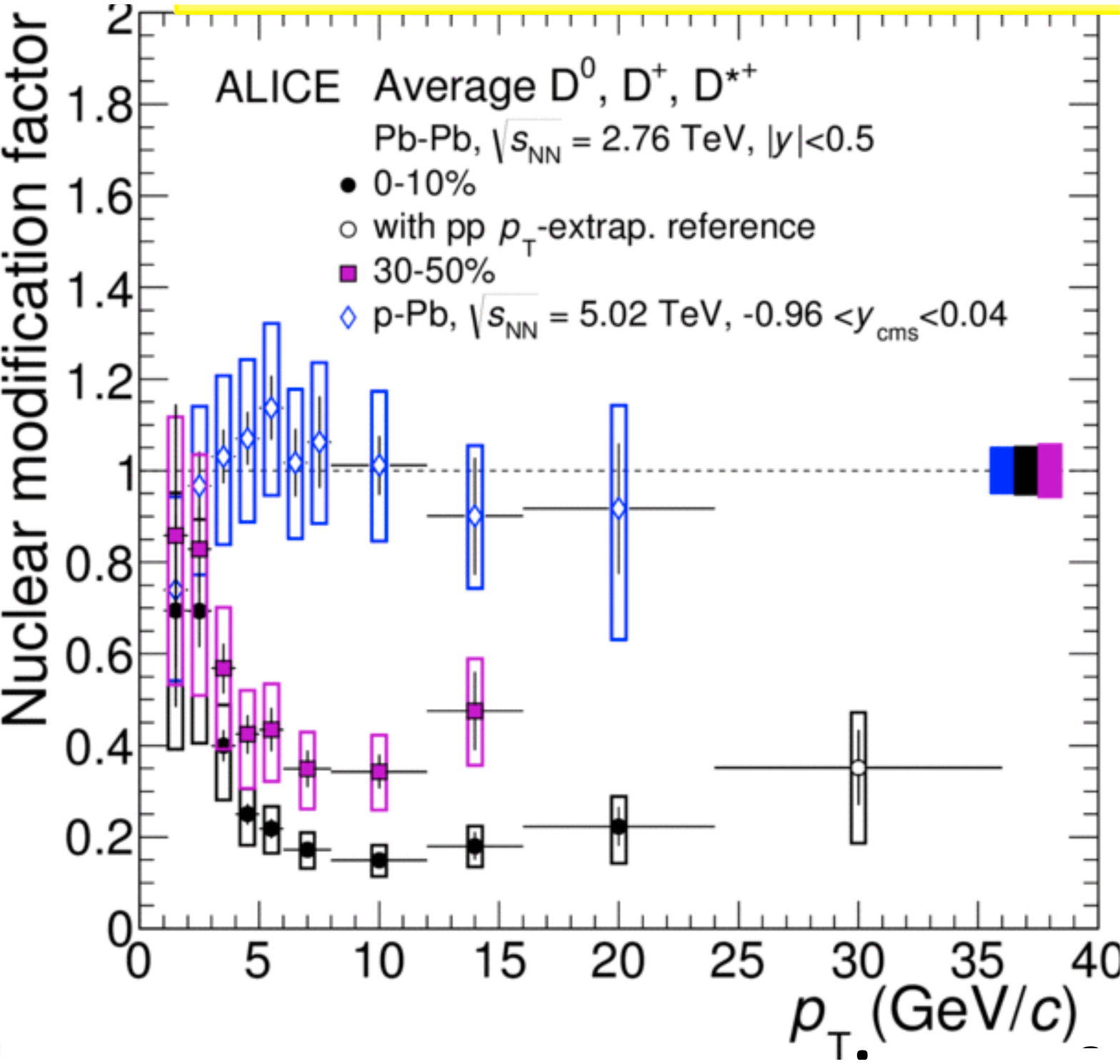}
\hskip0.05in
\includegraphics[width=0.32\textwidth]{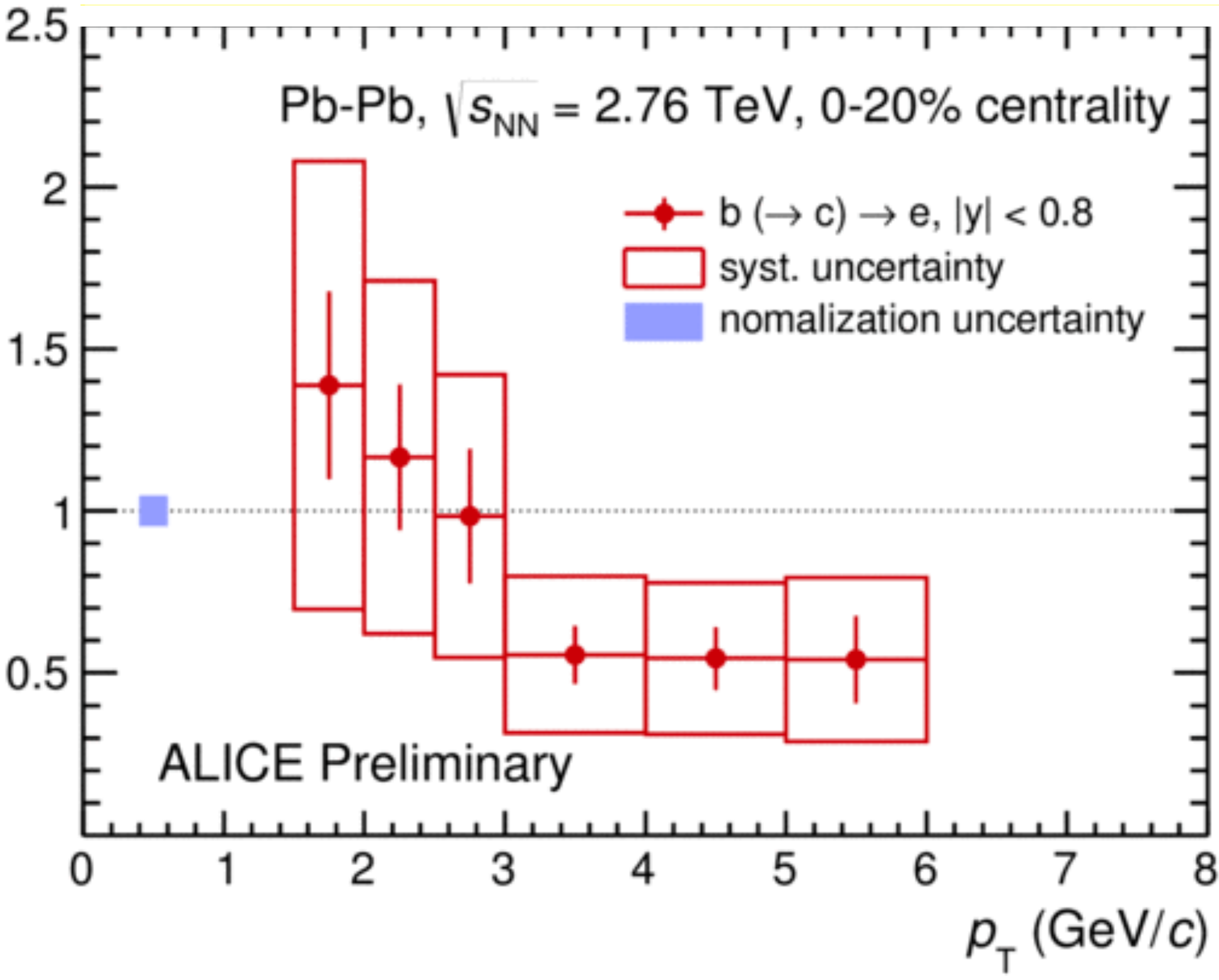}
\hskip0.05in
\includegraphics[width=0.32\textwidth]{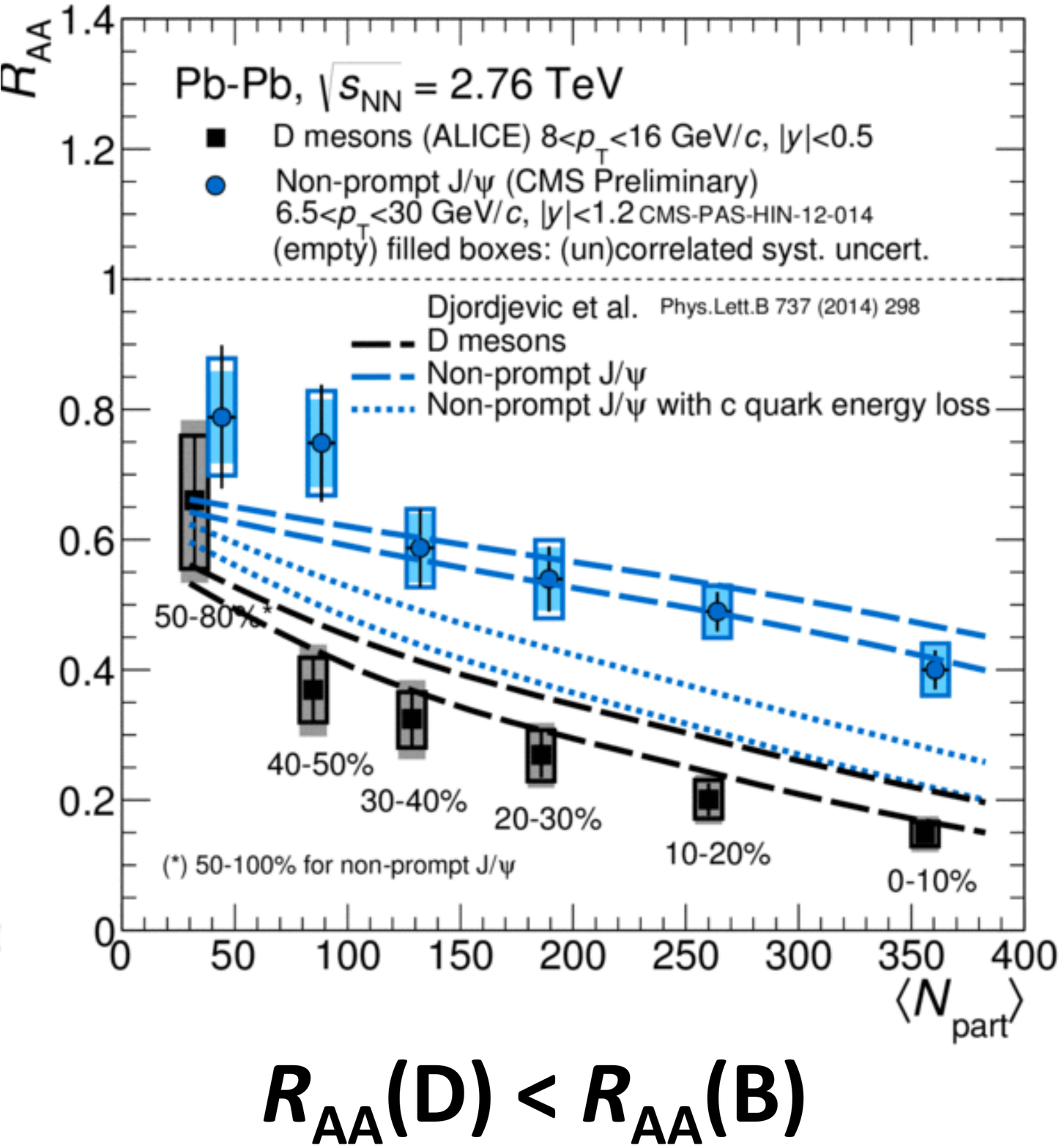}
}
\end{center}
\vskip-0.50in
\caption{
Left: $R^D_{AA}$ measured by ALICE for $D^0$ and $D^{(*)+}$ 
production in Pb-Pb and $p$-Pb collisions.
Middle: $R^B_{AA}$ measured by ALICE for high-impact-parameter
electrons in Pb-Pb collisions.
Right: $R^D_{AA}$ measured by ALICE as compared to 
$R^B_{AA}$ measured by CMS using non-prompt $J/\psi$ 
decays~\cite{hi_CMS}, in Pb-Pb collisions. Also shown
are theoretical predictions~\cite{suppression_theory}. }
\label{fig:suppression}
\end{figure}

Figure~\ref{fig:suppression}(left) shows $R^{}_{AA}$ plotted as a function
of $p^{}_T$ for $D^0$ and $D^{(*)+}$ production as measured by ALICE. The 
highest points correspond to $p$-Pb collisions and show little suppression;
presumably the incoming proton does not generate a sufficient QGP energy
density. The other points plotted correspond to Pb-Pb collisions and show 
large suppression. Figure~\ref{fig:suppression}(middle) shows $R^{}_{AA}$ 
for electrons in ALICE that have a large impact parameter with respect to
the primary interaction point, i.e., they originate from heavy flavor decays. 
This data also shows significant suppression. 
Figure~\ref{fig:suppression}(right) shows $R^{}_{AA}$ as measured by ALICE 
for $D$ mesons along with $R^{}_{AA}$ as measured by CMS~\cite{hi_CMS} 
for $J/\psi$ mesons having a large impact parameter with respect to the 
primary interaction; these $J/\psi$ decays originate from $B$ decays.
The plot shows that $R^D_{AA}<R^B_{AA}$, as expected due to
$m^{}_b > m^{}_c$. This data is an important confirmation of
this relationship. Also superimposed on the plot are several
theoretical predictions~\cite{suppression_theory}, which are 
consistent with the data.

\section{Heavy flavor decays}

The BESIII experiment (Ke) presented new measurements of branching 
fractions for the $\Lambda_c^+$ baryon decaying into a dozen hadronic 
final states; see Table~\ref{tab:besIII_hadronic}. With the exception 
of $B(\Lambda_c^+\ra p K^-\pi^+)$~\cite{Lambdac_Zupanc}, the BESIII 
results are the world's most precise and represent a significant 
improvement over previous results. BESIII also presented $\Lambda_c^+$ 
semileptonic branching fractions:
\begin{eqnarray*}
B(\Lambda_c^+\ra\Lambda \mu^+\nu^{}_\mu) & = & (3.49\,\pm 0.46\,\pm0.26)\% \\
B(\Lambda_c^+\ra\Lambda e^+\nu^{}_e) & = & (3.63\,\pm 0.38\,\pm0.20)\% \\
B(\Lambda \mu^+\nu^{}_\mu)/B(\Lambda e^+\nu^{}_e)
 & = & 0.96\,\pm 0.16\,\pm0.04\,.
\end{eqnarray*}
The last result constitutes a test of lepton universality in 
$\Lambda_c^+$ decays.

\begin{table}
\caption{\label{tab:besIII_hadronic}
$\Lambda^+_c$ branching fractions to hadronic final states, 
as measured by BESIII.} 
\begin{center}
\begin{tabular}{lccc}
\br
Mode & BESIII & 2014 PDG  & Belle \\
\mr
$p K^0_S$             & $1.52 \pm 0.08 \pm 0.03$ & $1.15 \pm 0.30$ & \\
$p K^-\pi^+$          & $5.84 \pm 0.27 \pm 0.23$ & $5.0  \pm 1.3$ & 
                                   $6.84 \pm 0.24 \,^{+0.21}_{-0.27}$ \\
$p K^0_S\pi^0$        & $1.87 \pm 0.13 \pm 0.05$ & $1.65 \pm 0.50$ & \\
$p K^0_S\pi^+\pi^-$   & $1.53 \pm 0.11 \pm 0.09$ & $1.30 \pm 0.35$ & \\
$p K^-\pi^+\pi^0$     & $4.53 \pm 0.23 \pm 0.30$ & $3.4  \pm 1.0$ & \\
$\Lambda \pi^+$       & $1.24 \pm 0.07 \pm 0.03$ & $1.07 \pm 0.28$ & \\
$\Lambda \pi^+ \pi^0$ & $7.01 \pm 0.37 \pm 0.19$ & $3.6  \pm 1.3$ & \\
$\Lambda \pi^+\pi^-\pi^+$ 
                      & $3.81 \pm 0.24 \pm 0.18$ & $2.6  \pm 0.7$ & \\
$\Sigma^0 \pi^+$      & $1.27 \pm 0.08 \pm 0.03$ & $1.05 \pm 0.28$ & \\
$\Sigma^+ \pi^0$      & $1.18 \pm 0.10 \pm 0.03$ & $1.00 \pm 0.34$ & \\
$\Sigma^+ \pi^+\pi^-$ & $4.25 \pm 0.24 \pm 0.20$ & $3.6  \pm 1.0$ & \\
$\Sigma^+ \omega$     & $1.56 \pm 0.20 \pm 0.07$ & $2.7  \pm 1.0$ & \\
\br
\end{tabular}
\end{center}
\end{table}

A higher precision test of lepton universality was presented by 
LHCb (Hamilton), which measured the ratio of branching fractions 
$R^{}_K\equiv B(B\ra K\mu^+\mu^-)/B(B\ra K e^+ e^-)$. Within the 
Standard Model (SM), this ratio is  within 0.1\% of unity~\cite{rkk_theory}. 
The measurement is challenging for LHCb due to the electrons. 
Photons reconstructed in the electromagnetic calorimeter that 
lie close to the candidate electron's trajectory are added to 
the electron's four-momentum to improve the resolution; as a 
consequence, the signal shape for the $K e^+e^-$ mass distribution must be
treated separately for one, two, or three ``recovered'' photons. In addition,
the background shape depends on whether the event was electron-triggered,
kaon-triggered, or passed some other trigger criterion; thus the different 
trigger streams are fitted separately. The result is 
$R^{}_K  = 0.745\,^{+0.090}_{-0.074}\,\pm0.036$ for $q^2< (6{\rm\ GeV})^2$,
which is $2.6\sigma$ below unity. This measurement has 
significantly higher precision than similar measurements made 
by Belle~\cite{rkk_belle} and BaBar~\cite{rkk_babar}.

\begin{figure}[htb]
\begin{center}
\includegraphics[width=0.68\textwidth]{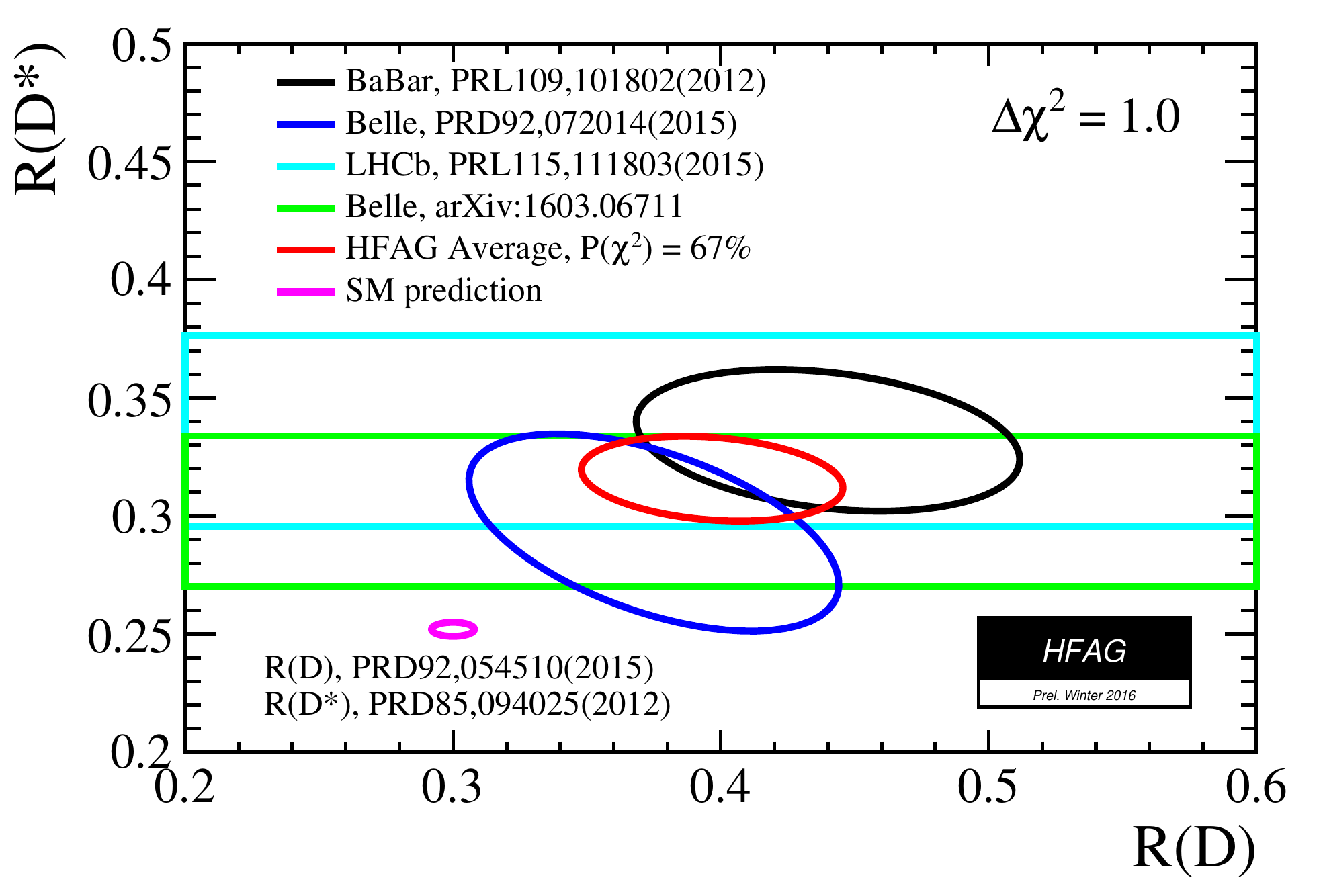}
\end{center}
\vskip-0.20in
\caption{
Measurements and theoretical predictions for $R(D^*)$ and $R(D)$, 
as compiled by the Heavy Flavor Averaging Group~\cite{rd_HFAG}. }
\label{fig:RD_hfag}
\end{figure}

The Belle experiment (King) presented a new result for the ratio
$R(D^{*})\equiv B(B\ra D^{*}\tau\nu^{}_\tau)/B(B\ra D^{*}\ell\nu)$ 
$(\ell= e,\mu)$ that uses ``leptonic tagging,'' i.e., events
in which the opposite-side $B$
is required to decay semileptonically, producing an electron or muon.
This requirement significantly reduces backgrounds. The analysis obtains
$231\pm 23$ signal ($D^*\tau\nu$) decays and $2800\pm57$ normalization
($D^*\ell\nu$) decays; the result is $R(D^*) = 0.302\pm 0.030\pm 0.011$, 
which, like a previous measurement by Belle using a hadronic tagging 
method~\cite{rdst_belle_hadronic_tag}, and also
an LHCb measurement~\cite{rdst_lhcb} (Hamilton), 
is significantly higher than the SM prediction 
of~0.25~\cite{rdst_theory}. Such a difference was also seen 
for the ratio $R(D) = B(B\ra D\tau\nu^{}_\tau)/B(B\ra D\ell\nu)$, 
i.e., the measured values were higher than the SM prediction. 
All measurements and the SM predictions are summarized in 
Fig.~\ref{fig:RD_hfag}.

\begin{figure}[htb]
\vskip-1.0in
\begin{center}
\includegraphics[width=0.55\textwidth]{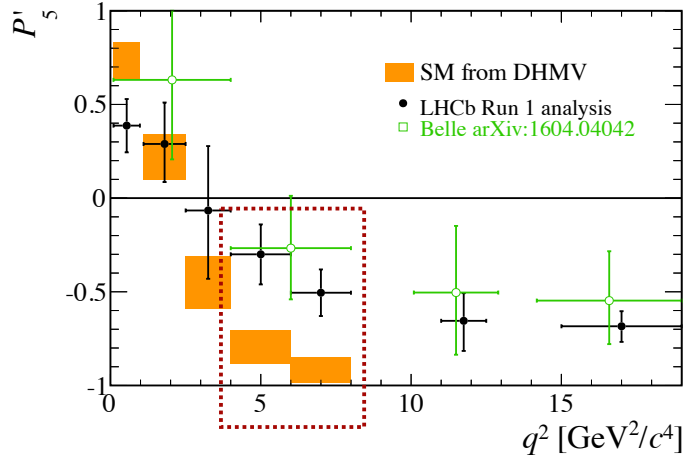}
\end{center}
\vskip-1.20in
\caption{
Measurements and theoretical predictions~\cite{kstmumu_theory2} 
for the parameter $P'_5$ measured in $B^0\ra K^{*0}\mu^+\mu^-$ decays.}
\label{fig:p5prime}
\end{figure}

Both Belle (King) and LHCb (Coutinho) presented results for 
the angular distribution of $B^0\ra K^{*0}\mu^+\mu^-$ decays. 
This distribution is parameterized as~\cite{kstmumu_lhcb}:
\begin{eqnarray}
\frac{d^4\Gamma}{dq^2 d\Omega} & \propto & 
\frac{3}{4}(1-F^{}_L)\sin^2\theta^{}_k + F^{}_L\cos^2\theta^{}_k 
* \frac{1}{4}(1-F^{}_L)\sin^2\theta^{}_k\cos 2\theta^{}_\ell
- F^{}_L\cos^2\theta^{}_k\cos 2\theta^{}_\ell + \nonumber \\
 & & S^{}_3\sin^2\theta^{}_k\sin^2\theta^{}_\ell\cos 2\phi + 
S^{}_4\sin 2\theta^{}_k\sin 2\theta^{}_\ell\cos \phi + 
S^{}_5\sin 2\theta^{}_k\sin \theta^{}_\ell\cos \phi + \nonumber \\
 & & \frac{4}{3}A^{}_{FB}\sin^2\theta^{}_k\cos\theta^{}_\ell + 
S^{}_7\sin 2\theta^{}_k\sin \theta^{}_\ell\sin\phi + 
S^{}_8\sin 2\theta^{}_k\sin 2\theta^{}_\ell\sin\phi + \nonumber \\ 
 & & S^{}_9\sin^2\theta^{}_k\sin^2\theta^{}_\ell\sin 2\phi
\end{eqnarray}
where $\theta^{}_k$ is the helicity angle of the $K^{0*}\ra K^+\pi^-$ decay, 
$\theta^{}_\ell$ is the helicity angle of the $\mu^+\mu^-$ system, 
and $\phi$ is the azimuthal angle between the $K^+\pi^-$ plane and
the $\mu^+\mu^-$ plane. There are eight underlying parameters, 
among which $F^{}_L$ is
the longitudinal polarization of the final state, and $A^{}_{FB}$ is 
the forward-backward asymmetry of the $\mu^+\mu^-$ system. Calculations 
of these parameters have large theoretical uncertainties, but
for the ratio $P'_5\equiv S^{}_5/\sqrt{F^{}_L(1-F^{}_L)}$ the leading
form factor uncertainties cancel~\cite{kstmumu_theory1}.
Figure~\ref{fig:p5prime} shows measurements of $P'_5$ from both 
Belle and LHCb as a function of $q^2$, which is the invariant 
mass squared of the $\mu^+\mu^-$ system. Superimposed on 
the data points are SM predictions~\cite{kstmumu_theory2}. 
There is good agreement between the measured values from the 
two experiments, but both experiments disagree with the SM 
prediction for $4{\rm\ GeV}^2/c^4 < q^2 < 8$~GeV$^2/c^4$. 
If these differences were statistical fluctuations, it is notable 
that both experiments observe fluctuations in the same direction 
for the same $q^2$ bins. More data from LHCb and the 
future Belle II experiment is needed to better 
understand this difference.

\section{Mixing and \cp\ violation}

The LHCb experiment (Carbone) presented three measurements
of $D^0$-$\dbar$ mixing and \cp\ violation: measurements of
$A^{}_\Gamma$ and $A^{}_{CP}$ in $D^0\ra K^+K^-/\pi^+\pi^-$ decays,
and a measurement of mixing in doubly Cabibbo-suppressed
$D^0\ra K^+\pi^-\pi^+\pi^-$ decays.

\begin{figure}[htb]
\vskip-0.70in
\begin{center}
\hbox{
\includegraphics[width=0.32\textwidth]{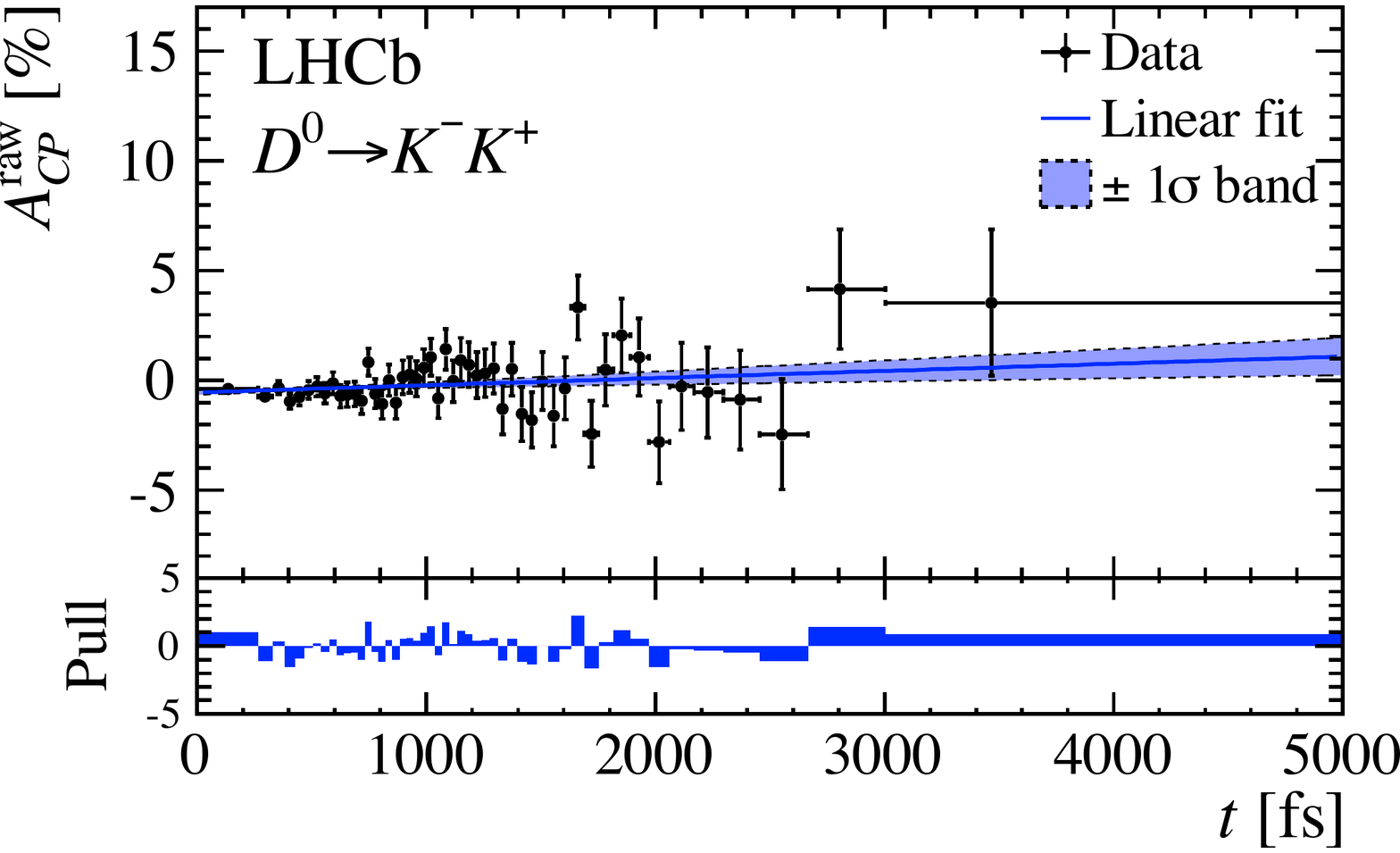}
\hskip0.05in
\includegraphics[width=0.32\textwidth]{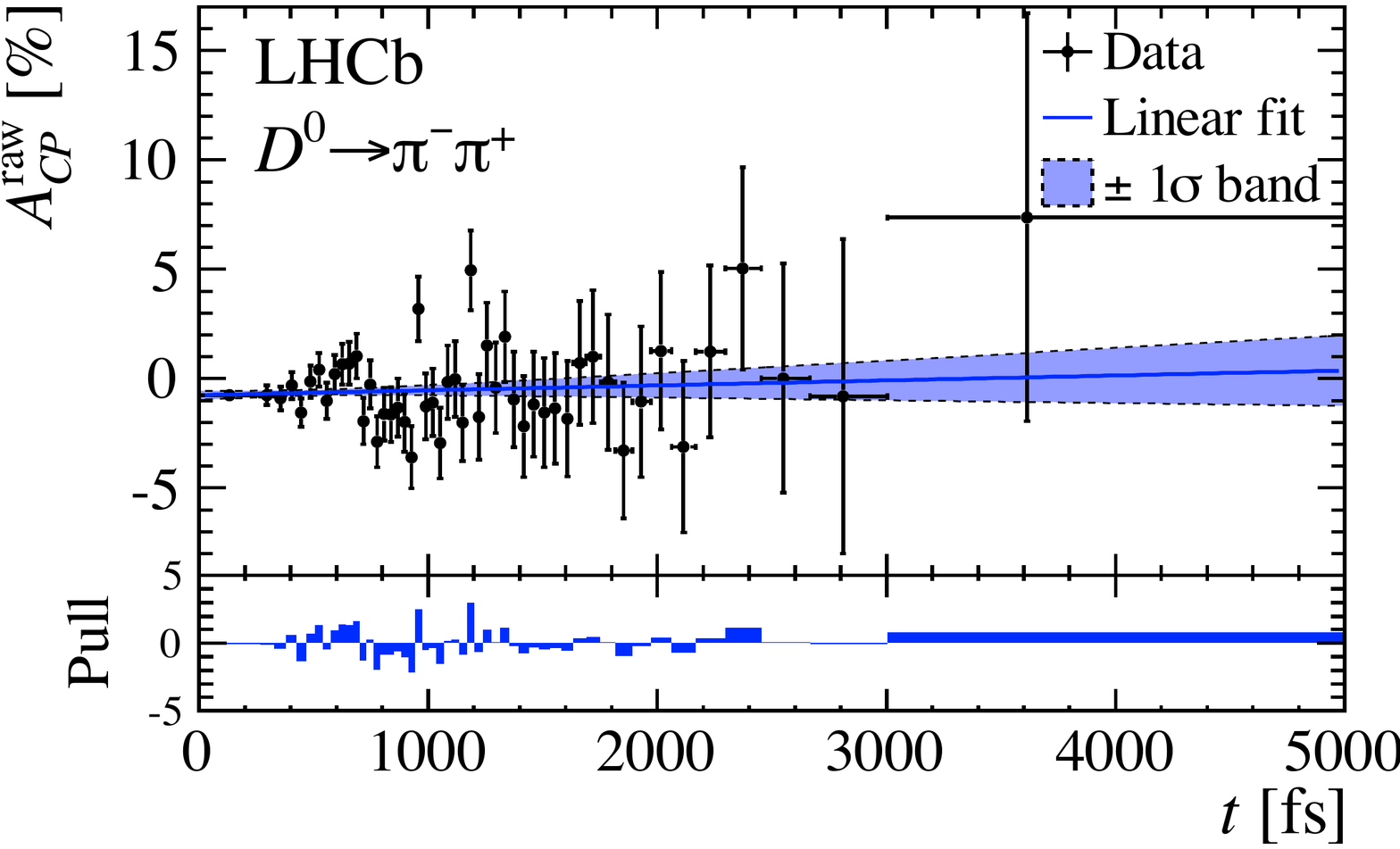}
\hskip0.05in
\includegraphics[width=0.32\textwidth]{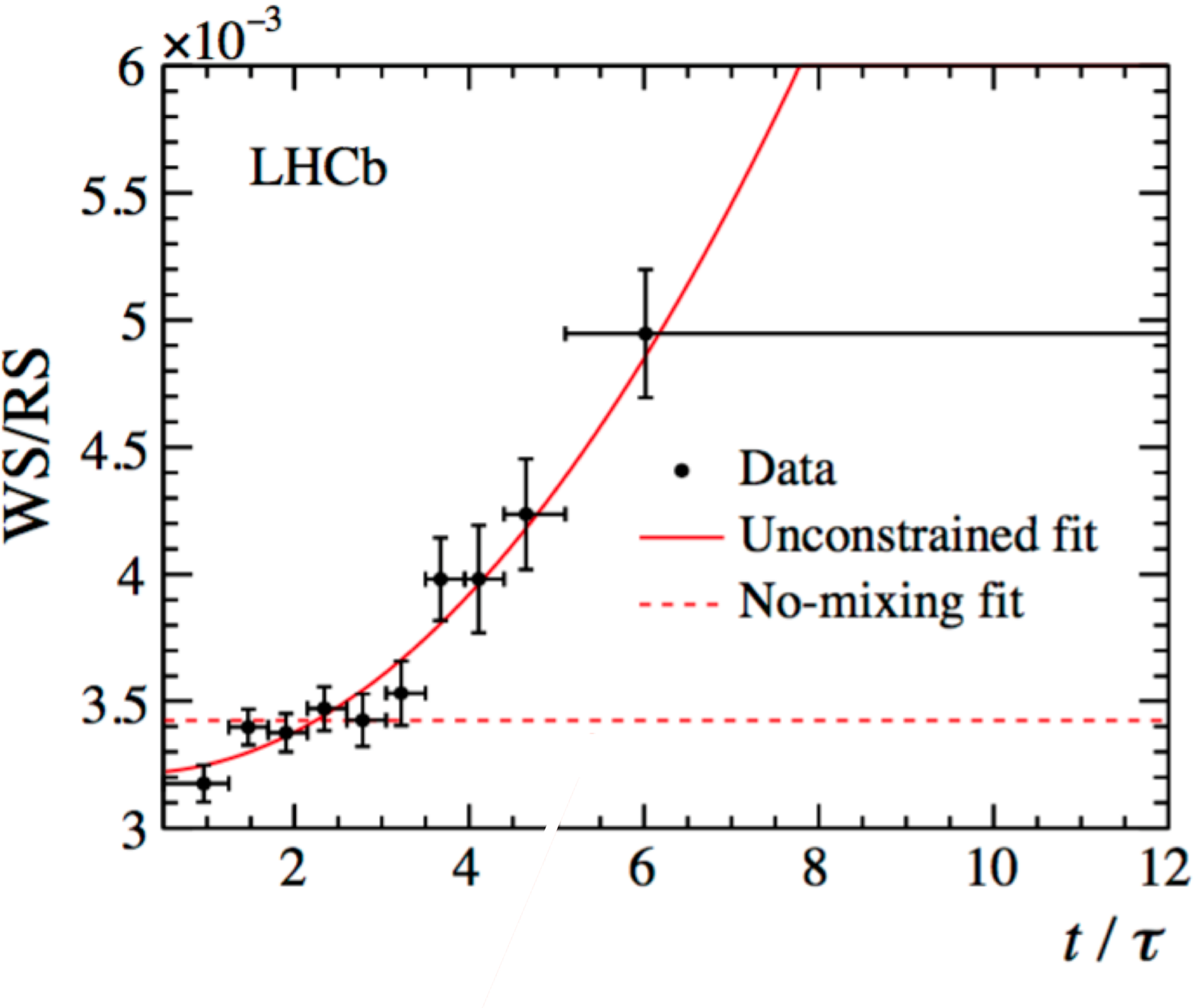}
}
\end{center}
\vskip-0.90in
\caption{
LHCb measurements of the 
\cp\ asymmetry in $D^0\ra K^+K^-$ decays (left);
the \cp\ asymmetry in $D^0\ra \pi^+\pi^-$ decays (middle); and
the decay time distribution of $D^0\ra K^+\pi^-\pi^+\pi^-$ decays (right). }
\label{fig:lhcb_cpasym}
\end{figure}

The parameter $A^{}_\Gamma$ is defined as the asymmetry in lifetimes
between $D^0$ and $\dbar$ decays:
$A^{}_\Gamma = (\tau^{}_{\dbar} - \tau^{}_{D^0})/(\tau^{}_{\dbar} + \tau^{}_{D^0})$.
One way to determine $A^{}_\Gamma$ is by fitting the decay
time distribution of the \cp\ asymmetry $A^{}_{CP}$ for
decays to a self-conjugate final state $f$:
\begin{eqnarray}
A^{}_{CP}(t) & = & \frac{dN(D^0\ra f)/dt - dN(\dbar\ra f)/dt}
{dN(D^0\ra f)/dt + dN(\dbar\ra f)/dt}\ 
\approx\ A^{\rm dir}_{CP} - A^{}_\Gamma\frac{t}{\tau}\,.
\end{eqnarray}
LHCb performs this measurement for $D^0\ra K^+K^-$ and $D^0\ra\pi^+\pi^-$
decays.  The flavor of the decaying $D^0$ or $\dbar$ is identified by 
reconstructing $D^{*+}\ra D^0\pi^+$ decays. The resulting $A^{}_{CP}$ 
distributions are shown in Fig.~\ref{fig:lhcb_cpasym}.
Performing a simultaneous fit to the $K^+K^-$ and $\pi^+\pi^-$
distributions gives $A^{}_\Gamma = (-0.125 \pm 0.073)$\%, 
which is the world's most precise measurement.

Using time-{\it integrated\/} (rather than time-dependent) samples 
of $D^0\ra K^+K^-$ and $D^0\ra\pi^+\pi^-$ decays, LHCb measures 
\cp-violating parameters $a_{CP}^{\rm ind}$ and 
$\Delta a_{CP}^{\rm dir} = a_{CP}^{\rm dir}(K^+K^-)-a_{CP}^{\rm dir}(\pi^+\pi^-)$.
For the analysis presented here the flavor of the $D^0$ is determined
by requiring that the $D^0$ originate from semileptonic 
$\bbar\ra D^0\mu^-\nu^{}_\mu$ decays; the charge of the accompanying 
$\mu^\pm$ then identifies the charm meson as $D^0$ or $\dbar$.
The results are $a_{CP}^{\rm ind}= (0.058\pm 0.044)\%$ and 
$\Delta a_{CP}^{\rm dir} = (-0.061\pm 0.076)$\%, which are 
both consistent with no \cp\ violation.

Finally, LHCb fits the decay time distribution of 
$D^0\ra K^+\pi^-\pi^+\pi^-$ decays to search for mixing in 
this doubly Cabibbo-suppressed decay mode.
The decay time distribution, normalized to that for the 
Cabibbo-favored decay $D^0\ra K^-\pi^+\pi^+\pi^-$, is given by
\begin{eqnarray}
\frac{dN}{dt} & \approx &  R^2_{K3\pi} \ -\ 
\kappa^{K3\pi} \cdot R^{}_{K3\pi} \cdot
y'_{K3\pi} \cdot \left(\frac{t}{\tau}\right) \ +\  
\frac{x^2+y^2}{4}\cdot \left(\frac{t}{\tau}\right)^2
\label{eqn:K3pi_decay}
\end{eqnarray}
where $R^2_{K3\pi}$ is the ratio of the $D^0\ra K^+\pi^-\pi^+\pi^-$ 
amplitude squared integrated over phase space to the 
$D^0\ra K^-\pi^+\pi^+\pi^-$ amplitude squared integrated 
over phase space; $\kappa^{K3\pi}$ is the coherence factor 
for $D^0\ra K^+\pi^-\pi^+\pi^-$ decays; and
$y'_{K3\pi} = y\cos\delta^{}_{K3\pi} - x\sin\delta^{}_{K3\pi}$, 
where $\delta^{}_{K3\pi}$ is the average strong phase difference 
between the $D^0\ra K^+\pi^-\pi^+\pi^-$ and $D^0\ra K^-\pi^+\pi^+\pi^-$ 
amplitudes.
The decay time distribution is shown in Fig.~\ref{fig:lhcb_cpasym}(right).
Fitting this distribution to Eq.~(\ref{eqn:K3pi_decay}) gives
$(x^2+y^2)/4 = (4.8\pm 1.8)\times 10^{-5}$, where the error includes 
systematic uncertainties. Due to a large correlation between the fitted 
terms $\kappa^{K3\pi} y'_{K3\pi}$ and $(x^2+y^2)/4$, the no-mixing hypothesis
$x=y=y'_{K3\pi}=0$ is rejected with a relatively high significance:~$8.2\sigma$.

\begin{figure}[htb]
\vskip-1.6in
\begin{center}
\vbox{
\includegraphics[width=0.70\textwidth]{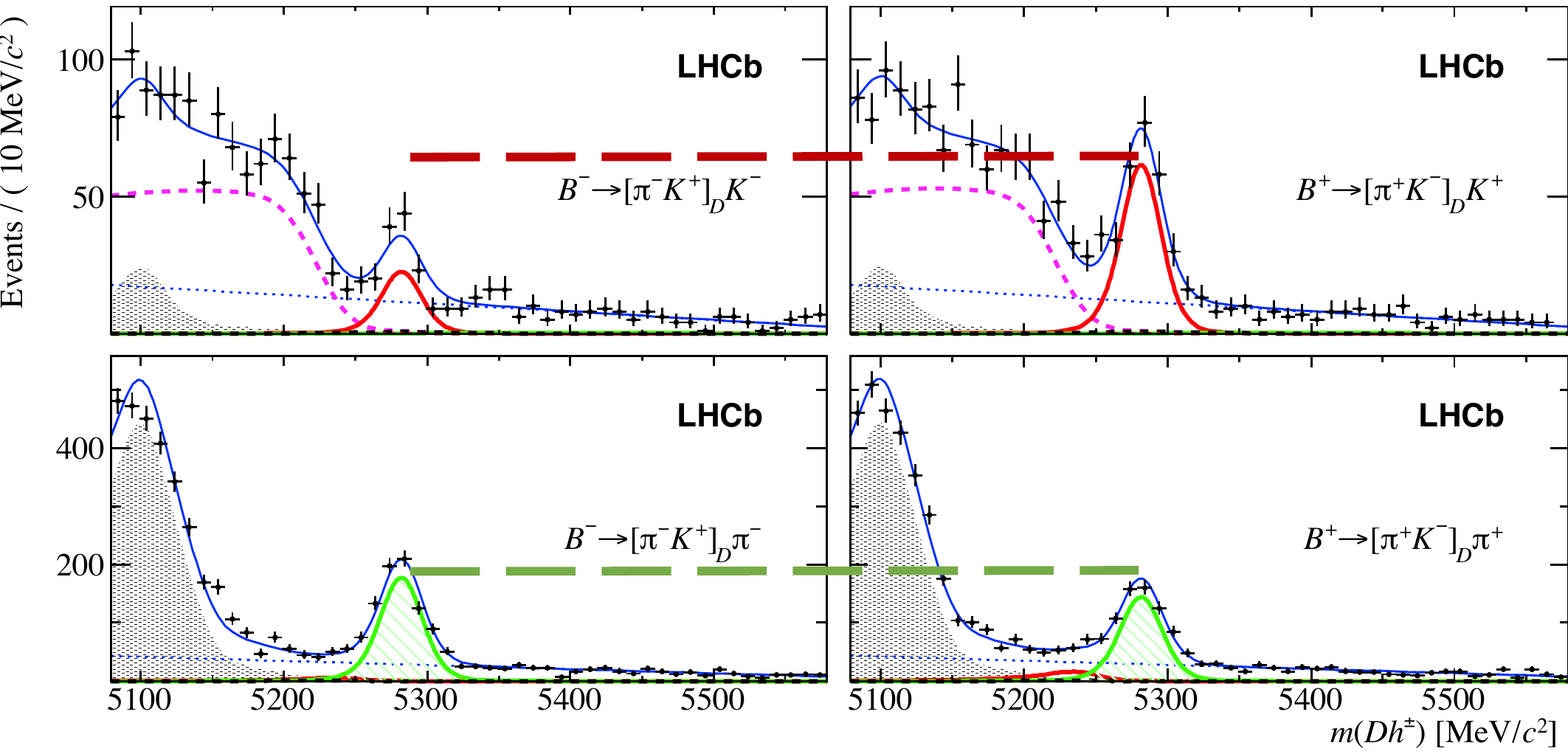}
\vskip-3.3in
\includegraphics[width=0.70\textwidth]{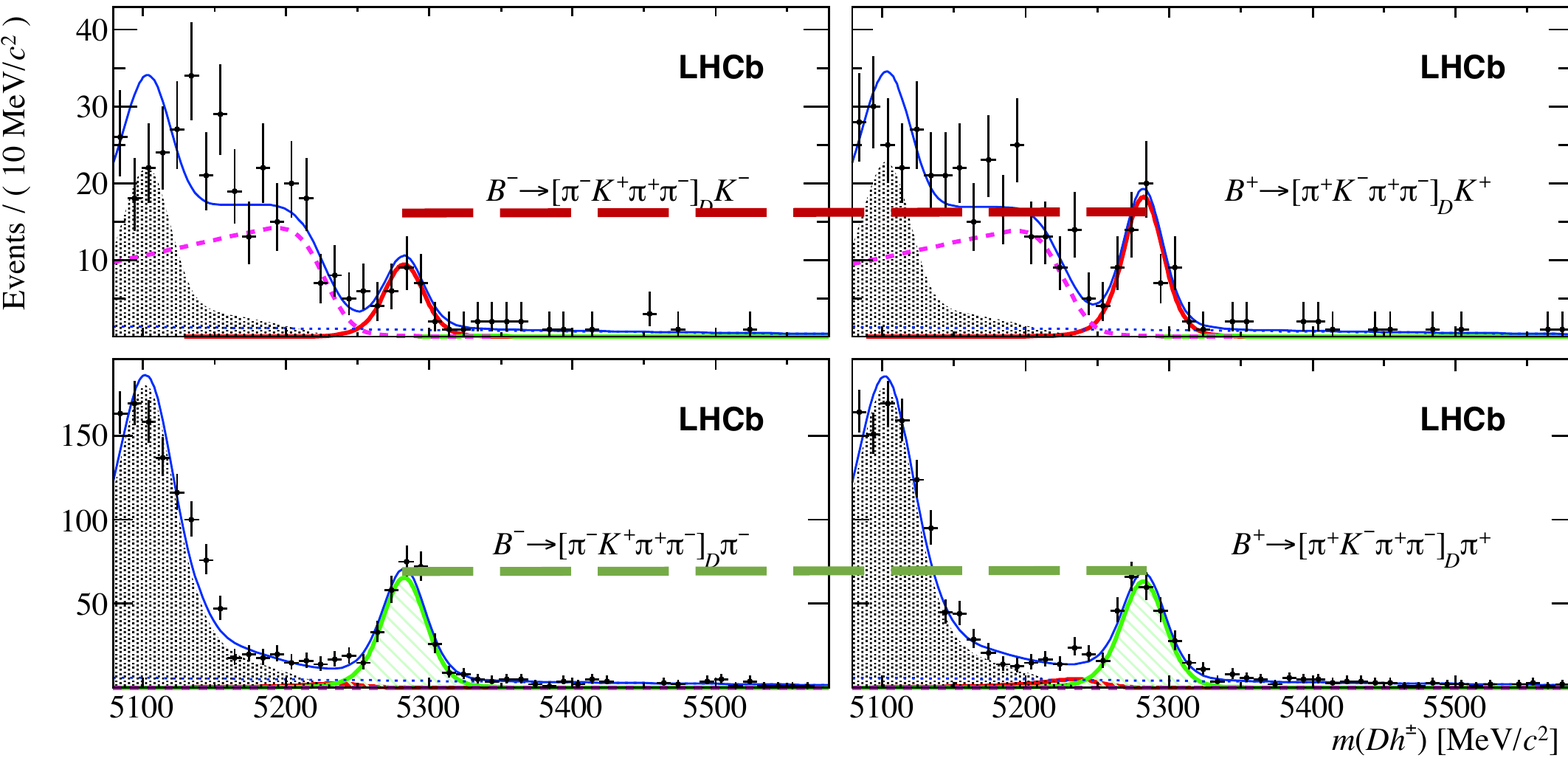}
}
\end{center}
\vskip-2.0in
\caption{
LHCb measurements of 
$B^-\ra (D^0,\dbar) K^-, (D^0,\dbar)\ra K^+\pi^-$ decays (top left) and 
$B^+\ra (D^0,\dbar) K^+, (D^0,\dbar)\ra K^-\pi^+$ decays (top right). A clear
\cp\ asymmetry is seen. The control samples $B^\mp\ra D^0\pi^\mp$ are also shown.
LHCb measurements of
$B^-\ra (D^0,\dbar) K^-, (D^0,\dbar)\ra K^+\pi^-\pi^+\pi^-$ decays (bottom left) and 
$B^+\ra (D^0,\dbar) K^+, (D^0,\dbar)\ra K^-\pi^+\pi^+\pi^-$ decays (bottom right). 
A clear \cp\ asymmetry is also seen, although the statistics are low.
The control samples $B^\mp\ra D^0\pi^\mp$ are also shown. }
\label{fig:lhcb_ads}
\end{figure}

LHCb (Whitehead) presented three measurements of the CKM phase 
$\phi^{}_3$ (or $\gamma$). The first measurement is based on the 
Atwood-Dunietz-Soni method~\cite{ADS}, in which one compares the 
rate for $B^+\ra (D^0,\dbar) K^+, (D^0,\dbar)\ra K^-\pi^+$
with that for the charge-conjugate decay
$B^-\ra (D^0,\dbar) K^-, (D^0,\dbar)\ra K^+\pi^-$.
These decays proceed through the interference of two amplitudes: a
Cabibbo-favored $B$ decay followed by a doubly Cabibbo-suppressed $D$ decay,
and a Cabibbo-suppressed $B$ decay followed by a Cabibbo-favored $D$ decay.
The phase difference between the overall $B^+$ and $B^-$ decay
amplitudes is $2\phi^{}_3$, which results in a difference in 
decay rates. Previous measurements by Belle~\cite{ADS_belle} 
and BaBar~\cite{ADS_babar} had relatively low statistics.
The LHCb data is shown in Fig.~\ref{fig:lhcb_ads}(top).
Systematic uncertainties are determined by repeating the
measurement with control samples of 
$B^+\ra (D^0,\dbar)^{}_{[K^-\pi^+]} \pi^+$ and
$B^-\ra (D^0,\dbar)^{}_{[K^+\pi^-]} \pi^-$ decays,
which should exhibit no difference in decay rates.
A total signal yield of $553\pm 54$ events is obtained,
and a \cp\ asymmetry is clearly visible; the statistical significance
of the asymmetry is almost $8\sigma$. LHCb subsequently applied this  method to
four-body $D^0\ra K^+\pi^-\pi^+\pi^-$ decays and observed a similar \cp\ asymmetry;
see Fig.~\ref{fig:lhcb_ads}(bottom). However, the statistics is lower
($159\pm 17$ events) and the difference seen is not statistically significant.
More data should establish a \cp\ asymmetry in this mode also.

The ATLAS experiment (Barton) measured
the $B^0$-$\bbar$ mixing parameter
$\Delta\Gamma^{}_d/\Gamma^{}_d$, where $\Delta\Gamma^{}_d$ is
the difference in decay widths between the two mass eigenstates, and
$\Gamma^{}_d$ is the mean decay width. For this measurement ATLAS
reconstructs $B^0\ra J/\psi K^0_S$ decays and fits the decay time
distribution. This distribution contains four terms:
\begin{eqnarray}
\frac{dN}{dt} & \propto & e^{-\Gamma t}
\left[
\cosh \frac{\Delta\Gamma t}{2} +
A^{}_P A^{\rm dir}_{CP}\cos(\Delta M t) + 
A^{}_{\Delta\Gamma}\sinh\frac{\Delta\Gamma t}{2} +
A^{}_P A^{\rm indir}_{CP}\sin(\Delta M t)\right]
\label{eqn:delta_gamma}
\end{eqnarray}
where $A^{}_P$ is the production asymmetry between $B^0$
and $\bbar$ mesons. This parameter
is measured by fitting the decay time distribution of 
flavor-specific $B^0\ra J/\psi\,K^{*0}, K^{*0}\ra K^+\pi^-$ 
decays, which should be purely exponential. The result
is $A^{}_P = (0.25\pm 0.48\pm 0.05)\%$. Inserting this into
Eq.~(\ref{eqn:delta_gamma}) along with the theoretical
values $A^{\rm dir}_{CP} = 0$, $A^{}_{\Delta\Gamma} = \cos(2\phi^{}_1)$, 
and $A^{\rm mix}_{CP} = -\sin(2\phi^{}_1)$, 
ATLAS obtains
$\Delta\Gamma^{}_d/\Gamma^{}_d = (-0.1\pm 1.1\pm 0.9)\%$.
This result is (surprisingly) more precise than measurements 
made at the $B$ factories and at LHCb, eclipsing the world's 
previous best measurement of $(1.7\pm 1.8\pm 1.1)\%$ (from 
Belle~\cite{deltagamma_belle}). It is consistent with the 
SM prediction of $(0.042\pm0.008)$\%~\cite{deltagamma_theory}.

\section{Spectroscopy}

There were numerous talks on spectroscopy; here we discuss two recent results.

\begin{figure}[htb]
\vskip-0.80in
\begin{center}
\hbox{
\includegraphics[width=0.48\textwidth]{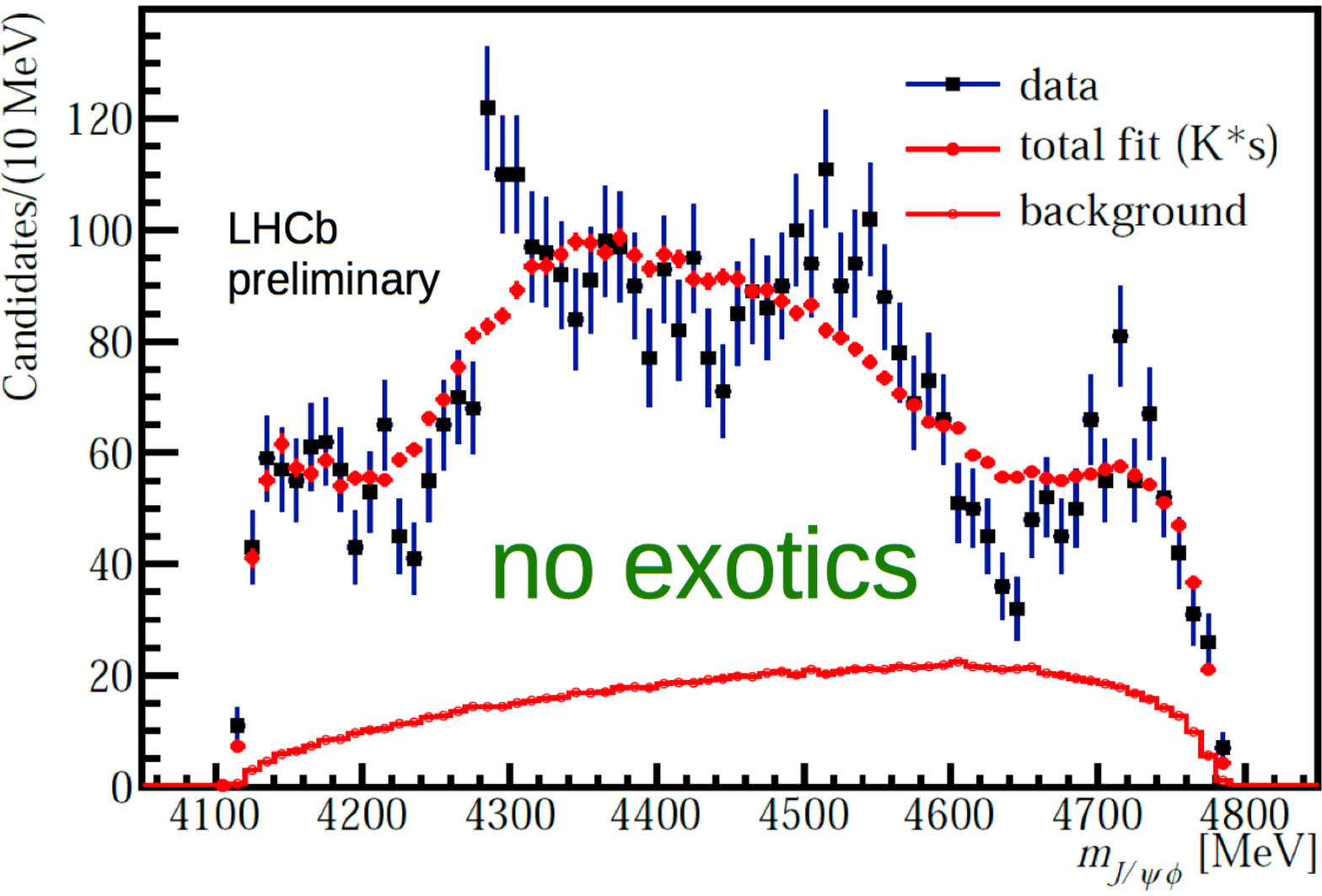}
\hskip0.10in
\includegraphics[width=0.48\textwidth]{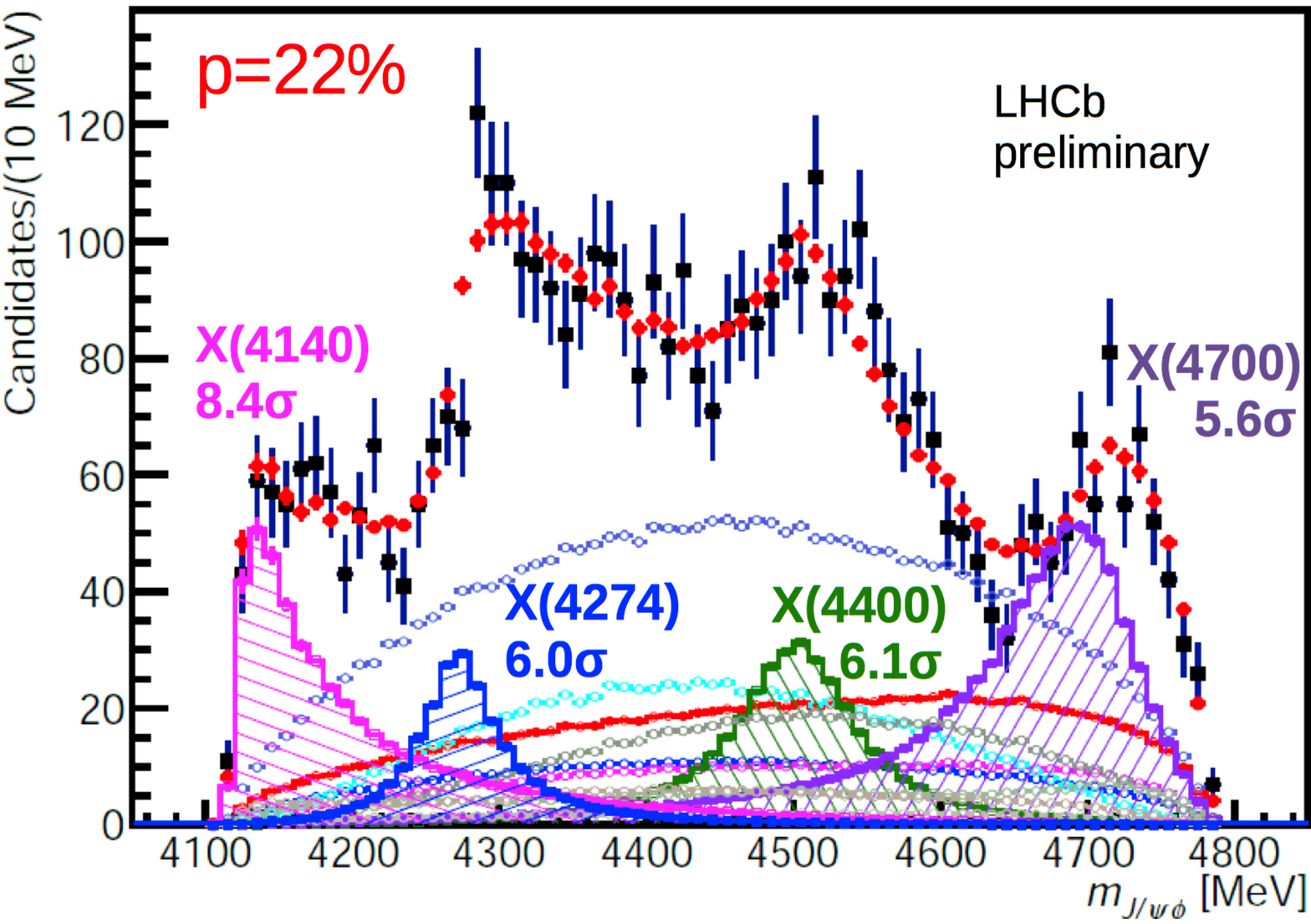}
}
\end{center}
\vskip-1.20in
\caption{
LHCb sample of $B^+\ra J/\psi\,\phi K^+$ decays.
The $M(J/\psi\,\phi)$ invariant mass distribution is fitted 
without (left) and with (right) states $X(4140)$, $X(4274)$, 
$X(4500)$, and $X(4700)$.}
\label{fig:lhcb_jpsiphi}
\end{figure}

The LHCb experiment (Dey) reconstructed a large sample of 
$B^+\ra J/\psi\,\phi K^+$ decays and studied the $J/\psi$-$\phi$ 
invariant mass distribution for unusual structure. This distribution is 
shown in Fig.~\ref{fig:lhcb_jpsiphi}(left) and exhibits four prominent peaks
corresponding to the known states $X(4140)$, $X(4274)$, $X(4500)$, and
$X(4700)$. Aside from these states no additional structure is apparent.
Fitting the $M(J/\psi\,\phi)$ distribution with these states plus background
gives a satisfactory goodness-of-fit: the $p$-value is~0.22. 
The fit projection is shown in Fig.~\ref{fig:lhcb_jpsiphi}(right), 
and the fit results are listed in Table~\ref{tab:lhcb_jpsiphi}.

\begin{table}
\renewcommand{\arraystretch}{1.2}
\caption{\label{tab:lhcb_jpsiphi}
Fitted parameters of exotic states used by LHCb to fit the
$M(J/\psi\,\phi)$ invariant mass distribution of 
$B^+\ra J/\psi\,\phi K^+$ decays.}
\begin{center}
\begin{tabular}{lccccc}
\br
State & $J^{PC}$ & significance  & Mass & Width & Fit fraction \\
\mr
X(4140) & $1^{++}$ & 8.4 & $4165\pm 4.5\,^{+4.6}_{-2.8}$ & 
$83\pm 21\,^{+21}_{-14}$ & $13.0\pm 3.2\,^{+4.8}_{-2.0}$ \\
X(4274) & $1^{++}$ & 6.0 & $4273.3\pm 8.3\,^{+17.2}_{-3.6}$ & 
$56\pm 11\,^{+8}_{-11}$ & $7.1\pm 2.5\,^{+3.5}_{-2.4}$ \\
X(4500) & $0^{++}$ & 6.1 & $4506\pm 11\,^{+12}_{-15}$ & 
$92\pm 21\,^{+21}_{-20}$ & $6.6\pm 2.4\,^{+2.5}_{-2.3}$ \\
X(4700) & $0^{++}$ & 6.1 & $4704\pm 10\,^{+14}_{-24}$ & 
$120\pm 31\,^{+42}_{-33}$ & $12\pm 5\,^{+9}_{-5}$ \\
\br
\end{tabular}
\end{center}
\end{table}

\begin{figure}[htb]
\vskip-0.45in
\begin{center}
\hbox{
\includegraphics[width=0.42\textwidth]{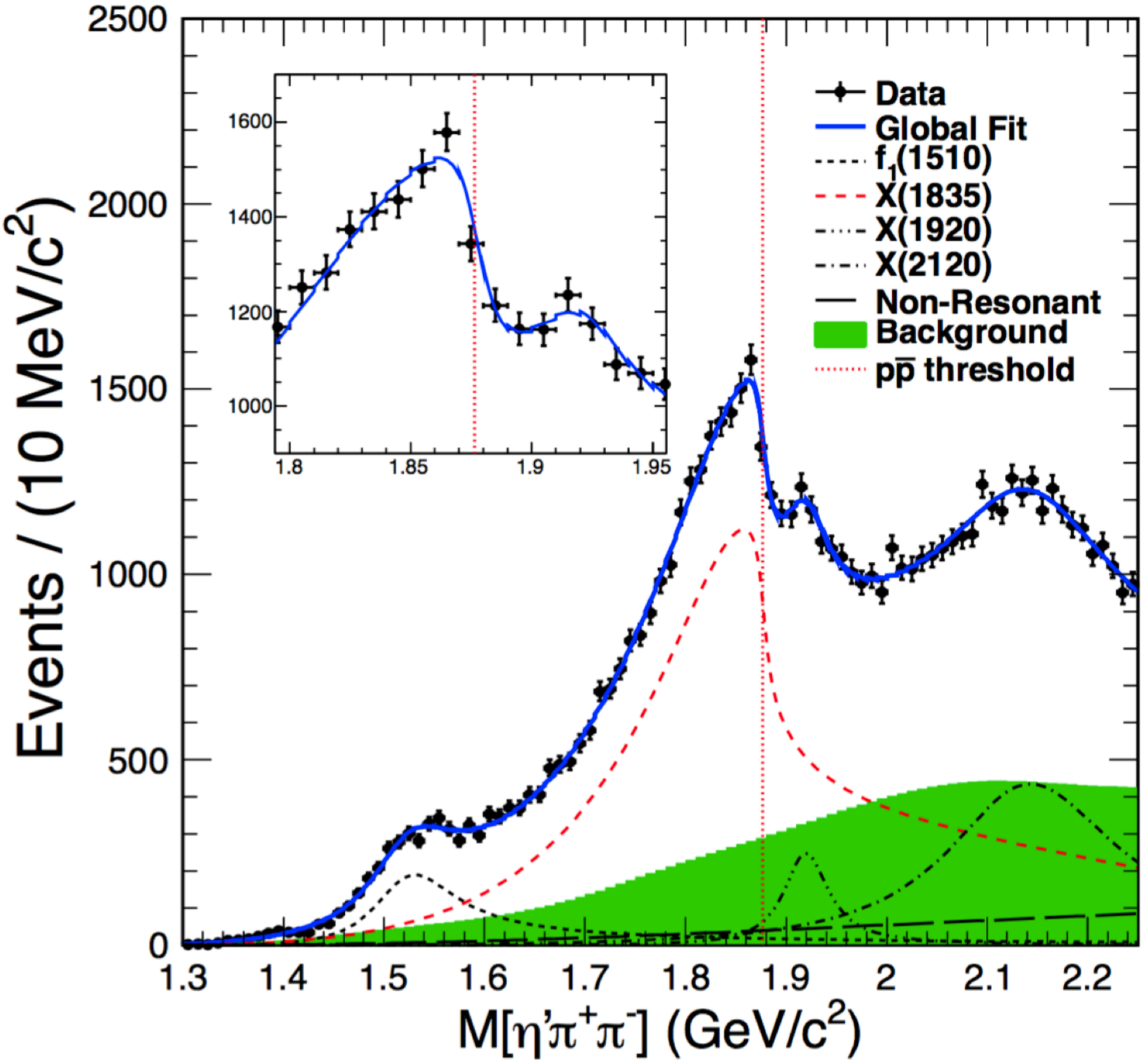}
\hskip0.30in
\includegraphics[width=0.42\textwidth]{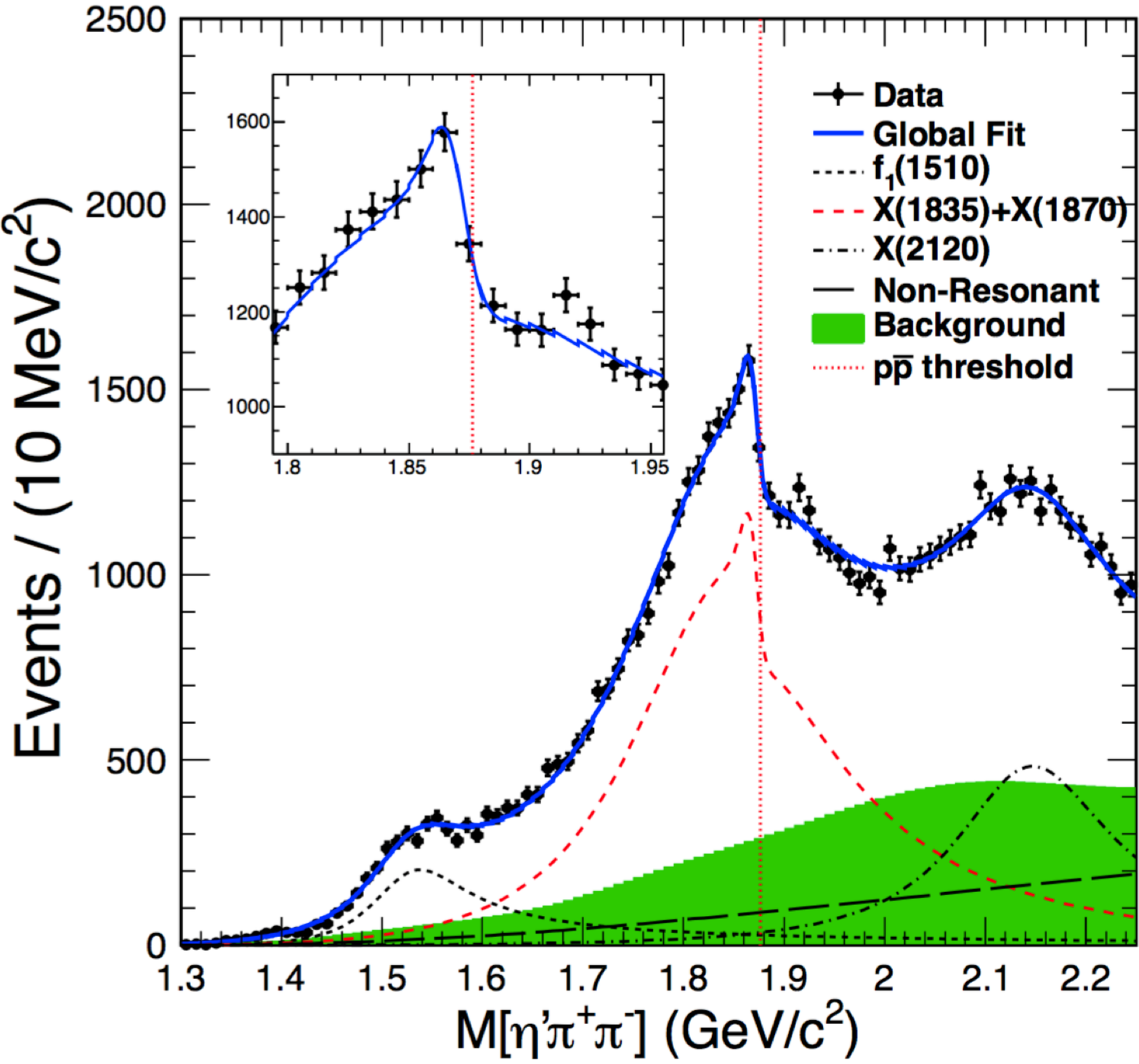}
}
\end{center}
\vskip-0.85in
\caption{
BESIII sample of $J/\psi\ra \eta'\pi^+\pi^-\gamma$ decays.
Left: fit result with a broad $X(1835)$ state and a narrow $X(1920)$ state.
Right: fit result with a broad $X(1835)$ state and a narrow $X(1870)$ state.
Both fits are satisfactory. }
\label{fig:bes3_1835}
\end{figure}

The BESIII experiment (Pelizaus) reconstructed an especially large 
sample of $J/\psi\ra \eta'\pi^+\pi^-\gamma$ decays with the goal of
identifying intermediate $J/\psi\ra X(1835)\gamma$ decays 
followed by $X(1835)\ra\eta'\pi^+\pi^-$.
The $\eta'$ is reconstructed in both $\eta'\ra\rho^0\gamma$ and 
$\eta'\ra\eta\pi^+\pi^-$ modes, where $\eta\ra\gamma\gamma$.
The resulting $M(\eta'\pi^+\pi^-)$ invariant mass distribution shows 
a clear peak near $M\approx 1835$~MeV/$c^2$, as expected, but it also 
shows a sharp drop at the $p\bar{p}$ threshold, which was unexpected.
The observed lineshape including this drop is subsequently modeled in two ways:
with a broad $X(1835)$ state and a narrow $X(1920)$ state (the latter being
just above $p\bar{p}$ threshold); and 
with a broad $X(1835)$ state and a narrow $X(1870)$ state (the latter being
just below $p\bar{p}$ threshold). Both parameterizations give satisfactory
fits, which are shown in Fig.~\ref{fig:bes3_1835}.

\section*{Acknowledgments}
The authors thank the BEACH 2016 organizers for a well-run workshop and
excellent hospitality. This research is supported by the U.S.\ Department 
of Energy.


\begin{thebibliography}{99}
\bibitem{charge-conjugates} 
Charge-conjugate modes are implicitly included unless noted otherwise
\end{thebibliography}
\end{document}